# DIAGNOSTIC IMAGE QUALITY ASSESSMENT AND CLASSIFICATION IN MEDICAL IMAGING: OPPORTUNITIES AND CHALLENGES

*Jeffrey Ma[1,2], Ukash Nakarmi[2], Cedric Yue Sik Kin[2], Christopher Sandino[3], Joseph Y. Cheng[2], Ali B. Syed[2], Peter Wei[2], John M. Pauly[3], Shreyas Vasanawala[2]*

[1] Department of Computing and Mathematical Sciences, California Institute of Technology
[2] Department of Radiology, Stanford University
[3] Department of Electrical Engineering, Stanford University

## ABSTRACT

Magnetic Resonance Imaging (MRI) suffers from several artifacts, the most common of which are motion artifacts. These artifacts often yield images that are of non-diagnostic quality. To detect such artifacts, images are prospectively evaluated by experts for their diagnostic quality, which necessitates patient-revisits and rescans whenever non-diagnostic quality scans are encountered. This motivates the need to develop an automated framework capable of accessing medical image quality and detecting diagnostic and non-diagnostic images. In this paper, we explore several convolutional neural network-based frameworks for medical image quality assessment and investigate several challenges therein.

## 1. INTRODUCTION

Magnetic resonance imaging (MRI) is an extensively used imaging modality in radiology. Unfortunately, MRI suffers from long data acquisition times and complex scanning protocols and parameters. Coupled with long acquisition times, physiological and patient motion, and complex scanning protocols, MR images often suffer from several motion artifacts, resulting in non-diagnostic quality images [1-3].

Current solutions to automatically detect and correct such artifacts are suboptimal. Despite the extensive training of MR technologists who operate scanners, these artifacts commonly go unrecognized. Thus, in some imaging centers, experienced radiologists review the images at the completion of the exam, which is an inefficient use of high-cost labor and capital equipment. In other centers, patients are simply asked to return for a repeat scan at a later date.

This necessitates an automated mechanism to detect image quality, which in one hand will save precious expert time and, on the other hand, will avoid patient-revisits/rescans due to non-diagnostic image quality. Recent data-based techniques, specifically Deep Learning (DL), present a promising framework for automating such detection and classification tasks [4-8]. Most widely, Convolutional Neural Network (CNN) architectures have been used to classify natural images that have distinct discriminants with very high accuracy, such as in the ImageNet competition [14]. However, detection and classification tasks in medical imaging are particularly different and challenging due to many reasons such as *i) the lack of clear discriminant between a diagnostic and non-diagnostic quality, resulting in subjectivity in classification*, *ii) unbalanced data sets*, and *iii) the lack of sufficient training data and reliable labels*. The consequence of such typical challenges in detection and classification tasks in medical imaging [5,18,19] is reflected in several recent works reporting comparatively low accuracies compared to accuracies in classifying natural images [14].

While, several works [5,18,19] focus on building more sophisticated models by integrating domain-specific knowledge and fine-tuning parameters, in this work, we closely investigate the challenges in the detection and classification task for medical image diagnostic quality assessment. We present our results on two DL models, compare their performance, and investigate several characteristics of the data set. The rest of the paper is organized as follows: section 2 provides details on our data and models; section 3 provides results in terms of standard performance metrics for two different DL models; section 4 covers several discussions and elaborations on challenges typical in detection and classification problem in medical imaging; and, finally, section 5 concludes the paper.

## 2. METHODS

### 2.1. Data and Preprocessing

We collected several 3D volume T2-weighted MR abdomen images at Stanford University and Lucile Packard Children's Hospital. The data consists of images from 69 volunteers with varying degrees of motion artifacts. Each patient data has about 30 slices, totaling to about 2111 512x512 MR images in the DICOM format. All data was acquired using the same imaging protocol and using a T2-weighted fast spin-echo sequence on 1.5T/3.0T MRI scanners. The entire dataset was rated by two different radiology experts into three classes of image quality: *0 – poor/non-diagnostic*; *1 – diagnostic*; *2 – excellent*. An example of each class is shown below in Figure 1.

We retrospectively redistributed the labeled dataset for two different classification tasks: I) For binary classification, images originally labeled as diagnostic (1) and excellent (2) were grouped into same class to give a new set of *Diagnostic quality class (1)* and *Non-diagnostic quality class (0)*; or II) A 3-class classification task into Non-Diagnostic (0), Diagnostic (1) and Excellent (2) as originally rated by the raters.

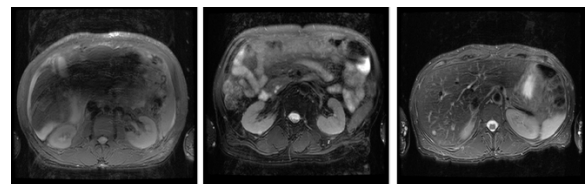

**Fig. 1:** An example of each of the three classes of image quality. Left-most is poor quality (0, non-diagnostic), the middle is diagnostic quality (1), right-most is excellent quality (2)

For each task, the dataset was split into training, evaluation, and testing sets using a 70-10-20 percent split, respectively. Images were normalized by dividing by the maximum intensity (255).

## 2.2. Model Architecture

In this paper we investigate the performance of two different CNN based DL architectures: *C1)* A simple 4-layer convolutional neural network and *C2)* A standard convolution kernel-based ResNet-10 architecture [15]. In CNNs, convolution kernels capture certain image features at each layer and are followed by a non-linear activation unit to enhance those features [16]. Typically, the deeper the network, the more non-linear features are captured by the network [20,21]. After analyzing image classification performance with comparatively deeper architectures such as ResNet-10 and ResNet-34, we observed that, for a comparatively simple classification task as in our case, where data is acquired using similar protocols, much non-linearity is not necessarily required. This is reflected in the activation maps from two different CNN based DL architectures, a simple 4-layer CNN and comparatively deeper ResNet-10, as shown in Figure 2 and discussed later.

Moreover, we believe that it is important to look at challenges specific to classification tasks in medical imaging as discussed in section 1. For this reason, we will focus our results and discussion on our 4-layered CNN architecture. Our architecture is as follows: an input greyscale MR image of size 512x512 flows through a 10x10 convolution, followed by a 7x7 convolution, and two 3x3 convolutions. Each convolutional layer is performed with stride 2, downsampling the image at each layer. After each convolutional layer, a Rectified Linear Unit (ReLU) activation is performed, followed by a batch normalization. The final feature tensor is flattened and fully-connected to 2 or 3 output logits, depending on whether the model is doing task I) binary classification or task II) 3-class classification. The architecture of the model is detailed in Figure 2.

The model was implemented on TensorFlow ver. 1.11.0. Each model was trained on the dataset for 10,000 steps with a batch size of 32 using the ADAM optimizer with a learning rate of 1e-3.

## 3. EXPERIMENTS AND RESULTS

### 3.1. Model Architecture Analysis

To understand image features extracted by CNN, we observed the path of an input image tensor as it flows through the model. We visualized activations for images from different classes to investigate discriminating image features learned by models for different classes. For this purpose, we created a set of toy example images, by taking a sample image from the diagnostic class and adding motion artifacts [22] to yield the same image with non-diagnostic quality. This toy example helped us keep all other image features identical except the ones resulting from motion artifacts, so that the discrimination between two classes is better visualized. Then, the activation maps in each of these two images allows us to observe if particular layer in model is correctly generalizing or differentiating between the classes.

We first built a baseline ResNet-10 model. Our analysis on the activation maps reveals that as we go deeper in the model, features learned by layers are not discriminating much between the diagnostic and non-diagnostic class as shown in Figure 3. It shows representative activation maps from the first layer (low-features) and last layer (deeper-features) of DL networks for our naïve 4-layer CNN and the deeper ResNet-10 model. We can clearly see that while low-level features (features from first layer) are discriminating between two classes in both models, the deeper features (features from last layer) from ResNet-10 are not very discriminating between two classes. This motivated us to focus more on a simpler 4-layer CNN network and put more effort towards investigating other classification challenges.

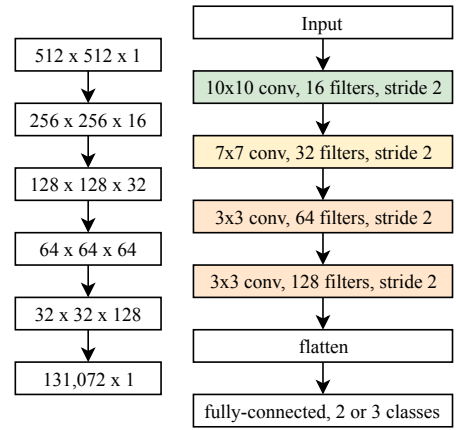

**Fig. 2**: A diagram depicting the 4-layer convolutional neural network (ConvNet) model. The shape of an image tensor throughout the model is shown on the left. The final layer is either 2 or 3 logits depending on if the model is being trained on the binarized dataset or the three-class dataset.

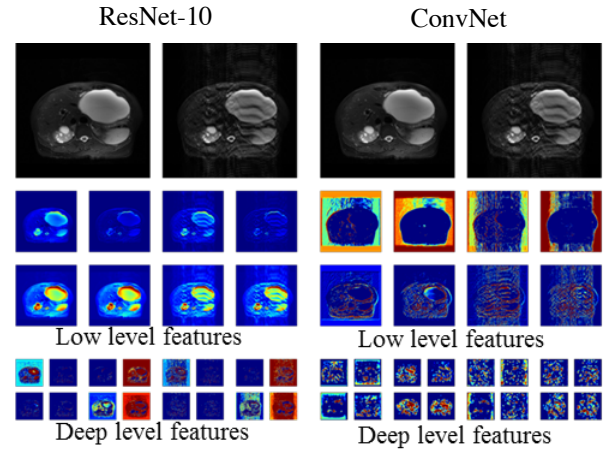

**Fig. 3**: Activation maps for images through the deep ResNet-10 model and the shallower ConvNet model. On the far left is an image of excellent quality, and on the middle-left is that image with heavy synthetic motion through the ResNet-10. On the right are the same two images through the ConvNet.

### 3.2. Model Performance and Accuracy

Figure 4 shows the performance of 4-layer CNN and ResNet-10 for both I) binary and II) 3-class classification tasks. We can see that our model achieves an 84% accuracy in the binary task and 65% in the 3-class classification task, which is much better than that of ResNet-10. Moreover, Figure 5 shows AUC-ROC curves that measure the performance of the model at various threshold settings and shows how well the model distinguishes between classes [14]. For both tasks, we can clearly see that the simple 4-layer CNN architecture performs better than ResNet-10 as depicted by the Area Under Curve (AUC) values for each case, further supporting our hypothesis that, for this problem, deeper and more sophisticated models are not necessarily required. However, we also note that for 3-class case, the performance gap between a simple CNN and the deeper ResNet-10 is much smaller, possibly due to the fact that as the number of classes increases, the level of discrimination between classes decreases. So, features from higher degrees of non-linearity (deeper layers) become more significant in classification than in tasks with fewer classes.

## 4. DISCUSSION

As discussed in section 1, classification tasks in medical images diagnostic quality assessment bear several challenges. Here we will present examples of such challenges that can be helpful in designing future data collection pipelines and in developing robust deep learning models for similar tasks.

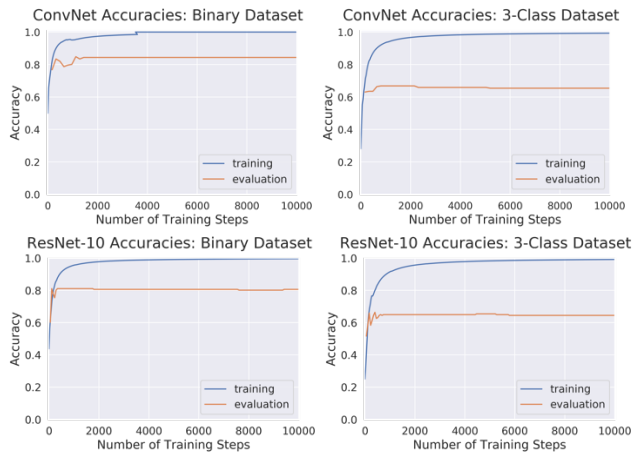

**Fig. 4**: Training and evaluation accuracy curves for ConvNet and ResNet-10 architectures on both the binary and 3-class dataset.

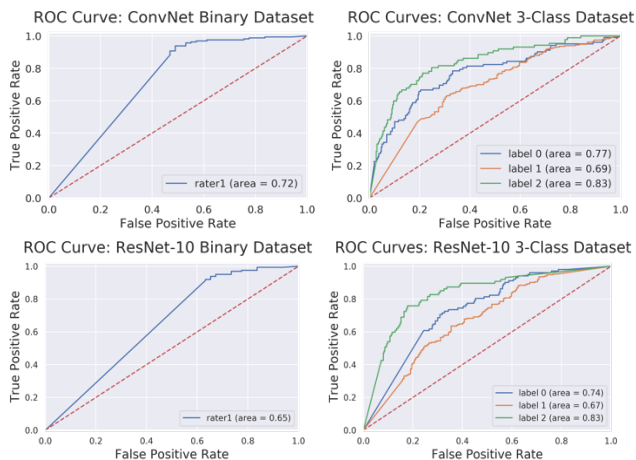

**Fig. 5**: AUC-ROC curves for the binary and three-class classification using ConvNet and ResNet-10 models. For reference, an excellent model has an AUC as close to 1 as possible, while a very poor model would have an AUC close to 0.5.

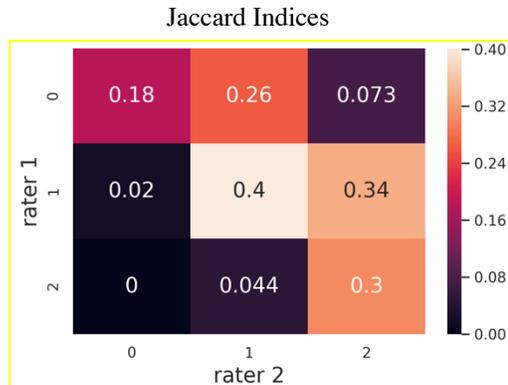

**Fig. 6**: Jaccard Indices matrix, which shows the Jaccard Index calculated for each pair of labels from each rater.

### 4.1. Inter-rater Subjectivity

We had the same dataset rated into 3 classes by two different radiology experts. Figure 6 shows a matrix of Jaccard indices [17], which measures the agreement between two raters for each class. Ideally, perfect consistency between raters would give a matrix of all diagonal elements equal to 1 and non-diagonal elements equal 0. However, we can see from Figure 6 that there exists a significant disagreement between the raters, suggesting subjectivity between raters in rating images for diagnostic quality.

This reveals the difficulty in determining the correct ground-truth labels for training data. To address this challenge, it is important to have the dataset labeled by multiple raters and agregate labels into a mean opinion score or another averaging metric so that the effect of subjectivity is insignificant.

### 4.2. Un-balanced Data

Unlike in natural image classification problems, where data distribution among each class is uniform, meaning that each class has comparatively the same numbers of samples, medical imaging data is non-uniformly distributed, which creates unbalanced data sets. For example, in our case, for the binary classification task, the data distribution contained 518 non-diagnostic images and 1592 diagnostic images, whereas, for 3-class classification, the data distribution consisted of 518 non-diagnostic, 1220 diagnostic, and 372 excellent images.

In such cases, the accuracy metric can be misleading, and the overall accuracy can be biased (highly influenced) by the accuracy of one class. To illustrate this, we present confusion matrices for both the binary and 3-class case in Figure 7. These matrices describe the performance of a classifier by comparing the predicted labels to the true labels of a dataset and identifying the classes that are most commonly misclassified and what they are misclassified as. We can clearly see that the accuracy is heavily influenced by the class that has highest numbers of samples (diagnostic class (1) in our case). Thus, it is important to have a balanced dataset and perform further analyses beyond studying evaluation accuracy, such as confusion matrix analysis and AUC-ROC curves.

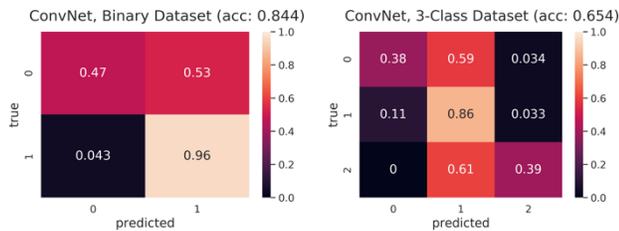

**Fig. 7**: Row-sum normalized confusion matrices for the binary and 3-class classification ConvNet models.

### 4.3. Unreliable labels

Because the classification task of image quality lacks clear discriminant between classes, the labels are often unreliable. For example, we show two cases below where a non-diagnostic quality image is labeled as diagnostic and vice-versa. So even if in inference time, the model correctly predicts the diagnostic quality class of an image (as shown in predicted classes below), its true label might not be reliable. These unreliable labels not only hinder the performance of model during the training phase but also affect the performance measurement during the inference time.

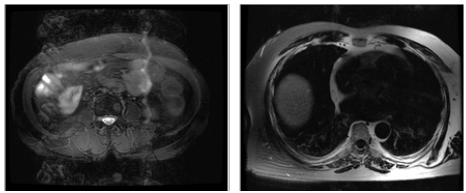

**Fig. 8**: Two images that have unreliable labels. The left image is a non-diagnostic image labeled as diagnostic, while the right image is diagnostic but labeled as non-diagnostic.

### 5. CONCLUSION

In this paper, we investigated several CNN based DL models for detecting artifacts in medical imaging and classify them in non-diagnostic and diagnostic quality. To summarize our result, our naive four-layer CNN architecture achieved an accuracy of 84% and AUC-ROC score of 0.79, for the binary classification task, indicating that a simple network was able to identify discriminants between non-diagnostic and diagnostic images. The same architecture achieved a three-class accuracy of 65%, largely due to the more subtle discriminants between diagnostic and excellent images. More importantly, we investigated and presented several challenges in the classification task-specific to medical imaging and its consequences on standard performance metrics used for classification tasks.

### 6. ACKNOWLEDGMENTS

This work is supported in part by the NIH R01EB009690 and NIH R01 EB026136 grants, and by GE Healthcare.